\documentclass[preprint,12pt]{aastex}
\usepackage[utf8]{inputenc} 
\usepackage[unicode]{hyperref}
\usepackage{nameref}
\usepackage{a4wide} 
\usepackage{amsmath,amssymb,amsfonts}
\usepackage{graphicx}
\usepackage{epstopdf}

\newcommand{\on}{N_{\mathrm{on}}}
\newcommand{\off}{N_{\mathrm{off}}}
\newcommand{\LM}{S_{\mathrm{LM}}}
\newcommand{\J}{HESS~J0632+057}

\begin{document}

\title{Time variability of the $\gamma$--ray binary HESS J0632+057} 

\author{S.~\v Stef\'anik, D.~Nosek}

\affil{Institute of Particle and Nuclear Physics \\ 
Faculty of Mathematics and Physics, Charles University \\
V Holesovickach 2, 180 00 Prague 8, Czech Republic}


\begin{abstract}
We study changes in the $\gamma$--ray intensity at very high energies observed from the $\gamma$--ray binary HESS J0632+057.
Publicly available data collected by Cherenkov telescopes were examined by means of a simple method utilizing solely the number of source and background events.
Our results point to time variability in signal from the selected object consistent with periodic modulation of the source intensity.
\end{abstract}


\section{Introduction}
Precise knowledge of variability in $\gamma$--ray fluxes observed from various emitters is considered essential for putting constraints on their intrinsic properties.
Among the most peculiar sources of such variable emission are the binary systems.
So far, only five $\gamma$--ray binaries have been detected~\citep{Paredes:2013}.
Out of these, four are known to emit very high energetic (VHE) $\gamma$--rays in the TeV energy domain.
All five binaries were recognized as high--mass systems emitting in X-ray band composed of a massive star of O or Be type and a compact object orbiting around it~\citep{Aliu:2014}.
Except for PSR~B1259--63 consisting out of an orbiting couple of Be star and a pulsar~\citep{Paredes:2013}, the nature of the compact object in other binaries, either a neutron star or a black hole, is unclear.
\par 
One of two most prominent acceleration mechanisms currently believed to take place in the binary systems assumes acceleration of charged particles in relativistic jets as a result of accretion onto a massive object like a black hole~\citep{Aliu:2014}.
The other scenario involves the outflow of relativistic particles from a pulsar.
Subsequent $\gamma$--ray emission processes may include synchrotron radiation and inverse Compton scattering of relativistic particles off the ambient photon population.
Studies of variability patterns of $\gamma$--ray binaries might provide valuable information on radiation processes in these systems.
Besides that, proper investigation of data is needed as the nature of the variability itself is also not evident.
This is due to the fact that the orbital motion of objects can cause a periodic modulation of the observed flux without any intrinsic variability of sources~\citep{Aliu:2014}.
\par 
Development of $\gamma$--astronomy in the past decade has brought a strong demand for reliable and precise statistical method for variability studies.
Various methods are used for investigation of temporal changes of $\gamma$--ray intensities observed by different instruments.
Usually, photon fluxes are determined first and their overall trend is then analysed.
Fit to a constant is often performed together with a subsequent check whether the residual sum of squares for fluxes is drawn from the $\chi^2$ distribution, see,
e.g.,~\citep{Aharonian:2005}, \citep{Acero:2010}, \citep{Abramowski:2010}.
Variability index is also used to examine changes in the lightcurves (e.g.~\citep{Aliu:2014}).
It is derived as the maximum likelihood ratio for different hypotheses, namely those of the flux constant throughout the whole observational period and of the flux optimized in each time bin~\citep{Nolan:2012}.
Other techniques, for example, Bayesian blocks--based method~\citep{Scargle:1998}, attempt to recognise change points in the lightcurves, for application see, e.g.,~\citep{Mayer:2013}.
The list of used methods is not exhaustive; however, it demonstrates the wide range of different approaches to the problem of time variability in $\gamma$--ray astrophysics.
\par
Our modification of the on--off method~\citep{Li:1983} aims to determine a level of significance for an excess or deficit of counts in individual measurements when compared to the reference source intensity previously ascertained from other observations~\citep{Nosek:2013}.
In the on--off method there is no demand for the calculation of the flux or other quantities.
It works only with the numbers of events detected in the on--source and reference off--source region provided that their exposures are known.
The method is equally suitable for any observations regardless of the experimental technique, thus making the comparison of data detected by different instruments possible.
It allows to check for intensity changes in the ranges of times, energies or zenith angles, for example, as long as reasonable estimates for the source intensity exist.
We use this technique to examine whether the $\gamma$--ray binary \J~exhibits any temporal changes in the numbers of $\gamma$--ray events measured from its direction.


\section{Method}
\label{sec:method}

The Li--Ma method is widely used in VHE $\gamma$--ray astronomy for determining a level of significance of a photon excess above background when validating the source presence in a given region \citep{Li:1983}.
In this method, one assumes the test of the null hypothesis stating that there is no source present in the investigated on--source region.
The on--source region encompasses the potential $\gamma$--ray emitter whereas the off--source region is considered to be free of point sources and thus suitable for the background estimation.
In order to account for different extent (e.g.~temporal or spatial) and unequal observing conditions of both regions, an on--off parameter $\alpha$, the ratio of the on-- and off--source exposures, is needed. 
\par 
A straightforward modification of this technique allows one to estimate the significance of an excess or deficit of the number of events when compared to the known source activity \citep{Nosek:2013}.
The modified on--off method assumes that the $\gamma$--ray emitter has already been positively identified in the potential hotspot.
A source parameter $\beta > 0$ is introduced to characterize its intensity.
Then the statement of the null hypothesis is that the source attains intensity of previously chosen value $\beta$, i.e.~$\on = \alpha \beta \off$~\citep{Nosek:2013}, where $\on$ and $\off$ are the numbers of events detected in the on-- and off--source regions, respectively.
\par 
Taking the source parameter $\beta$ being equal to unity one recovers the original no--source assumption.
Alternatively, the inequality \mbox{$\beta > 1$} expresses that an excess of counts above the source intensity will be tested while \mbox{$0 < \beta < 1$} implies the test of their deficit.
\par 
A level of significance for the rejection of the no--source assumption is given in terms of Li--Ma statistics \citep{Li:1983}.
A modification of the original significance formula (Eq.~17 in \citep{Li:1983}) for the assumption of the constant source intensity leads to a similar equation, the only difference being the transformation $\alpha \rightarrow \alpha \beta$ \citep{Nosek:2013}, i.e.
\vspace{-0.3cm}
\begin{equation}
\label{eq:lima}
\LM = s\sqrt{2}\left\lbrace \on \ln{\left[ \frac{(1+\alpha\beta)}{\alpha\beta} \frac{\on}{\on + \off} \right]} + \off \ln{\left[ (1+\alpha\beta) \frac{\off}{\on + \off} \right]} \right\rbrace^{\frac{1}{2}}.
\end{equation}
The $s$--term in Eq.~(\ref{eq:lima}) ($s = \pm 1$) accounts merely for the sign of the whole expression, depending whether an excess ($\LM>0$) or deficit ($\LM<0$) of events is observed.
\par 
Given a set of observations the Li--Ma statistics can be evaluated in each individual case for different values of the source parameter $\beta$ chosen in advance at one's discretion.
Based on the $\LM$ values, one can infer whether measured data disagree with the null hypotheses given by various benchmark values of $\beta$.
Since the source intensity is not given in advance the source parameter $\beta$, by which it is represented, is essentially free.
Thus, in addition to the test of the hypothesis of chosen source intensity, the Li--Ma statistics can be harnessed to derive confidence intervals from an observed set of data at a given level of significance.
Series of such confidence intervals provides a powerful tool for examination of the progress of the source $\gamma$--ray activity in time.


\section{Data analysis}
\label{sec:analysis}

We studied temporal evolution of the observed VHE $\gamma$--ray intensity from a $\gamma$--ray binary system \J~ using the modified on--off method.
In our calculations we used publicly available data obtained by experiments employing the Cherenkov telescopes.
In particular, we utilized those measurements where the number of detected on-- and off--source events, $\on$ and $\off$, and the on--off parameter $\alpha$ for different periods of time are presented.
Given the values of these quantities determined by the experiment the only remaining quantity left in Eq.~(\ref{eq:lima}) is the source intensity expressed by the source parameter $\beta$.
\par 
In the following sections, our results are visualized in plots showing a sequence of confidence intervals, $\langle \beta_{-}, \beta_{+} \rangle$, arranged in a chosen time order.
In these plots, non--overlapping confidence intervals are thought to represent exceptional measurements of source counts revealing emission or absorption features in the overall lightcurve.
For each triplet $(\on,~\off,~\alpha)$, confidence intervals for the source parameter $\beta$ were determined numerically such that the Li--Ma significance in Eq.~(\ref{eq:lima}) satisfies $|\LM(\on,~\off,~\alpha;~\beta)|<S_{\mathrm
{C}}$, where $S_{\mathrm{C}}$ is chosen critical value of a confidence level.
The span of confidence intervals is driven by the number of detected events.
Firmer restrictions on the source parameter are obtained for larger numbers of events.
We also determined benchmark estimates of the source parameter $\beta$ derived as average values of the ratio of observed and expected on--source events over individual time intervals, i.e.~$\beta = \langle \on / \alpha \off \rangle$.


\subsection{\J}

\begin{figure}[t!]
	\begin{center}
		\includegraphics[width=\columnwidth]{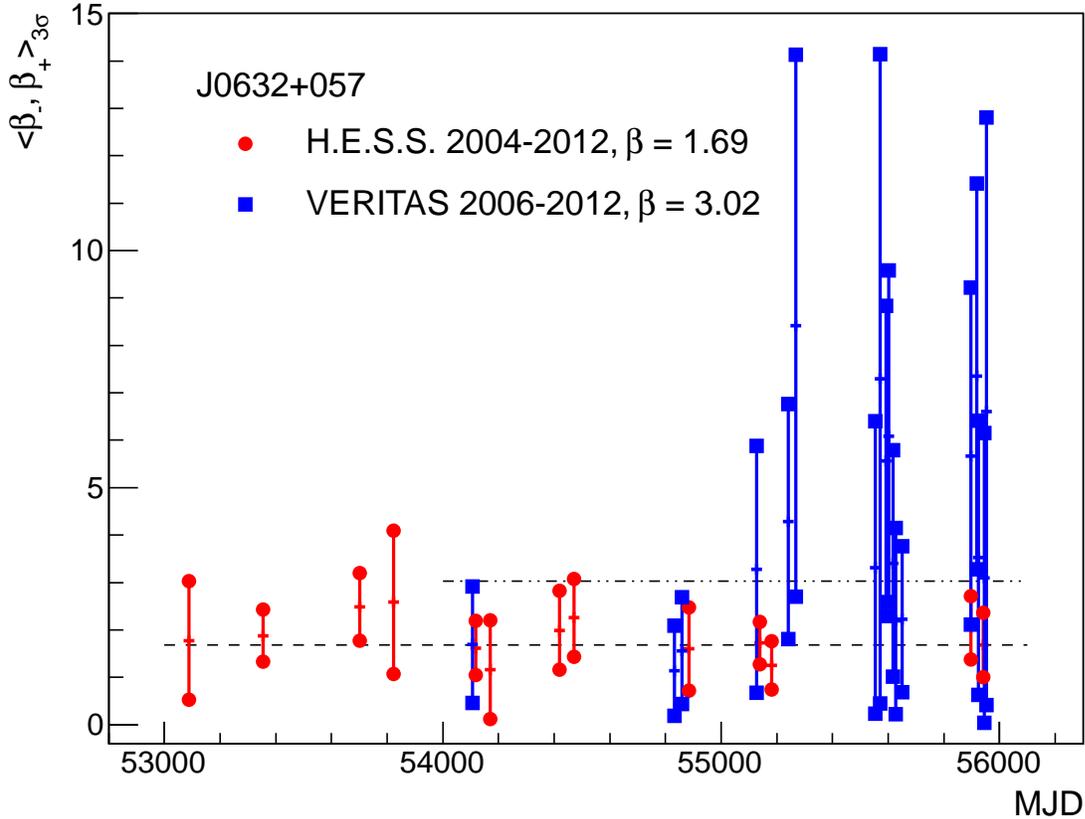}
		\caption{
			$99.7\%$ confidence intervals for the source parameter $\beta$ of the $\gamma$--ray binary \J~\protect\citep{Aliu:2014} are shown as a function of Modified Julian Date.
			Vertical lines with markers indicate the span of confidence intervals.
			The values of the source parameter $\beta = \on / \alpha \off$ derived from individual measurements are also indicated.
			The horizontal dashed and dotted lines denote the average source parameters $\beta_{\mathrm{HESS}}=1.69$ and $\beta_{\mathrm{VER}}=3.02$, respectively, calculated from the whole sets of data collected by the HESS and VERITAS collaborations. 
		}
		\label{fig:J0632_MJD}
	\end{center}
\end{figure}

\par
Gamma--ray binary \J~ was observed at very high energies by the VERITAS instrument during the 2006--2012 campaign~\citep{Aliu:2014}.
The numbers of events registered in individual measurements were less than thirty in the on--source region and in the range of several tens to two hundred in the off--source regions.
The source was detected above the energy threshold of $230$~GeV at a $15.5\sigma$ level of significance deduced from the total observational live time of 144 hours.
The $\gamma$--ray binary was also a target of the 2004--2012 observations of the HESS telescope system~\citep{Aliu:2014}.
With the exception of one observational period, all measurements yielded the numbers of detected events ranging from over ten to several hundred.
The HESS collaboration claimed detection of the source above the energy $\sim 220$~GeV at a $13.6\sigma$ level of significance inferred from 53.5 hours of live time observations.
Combined results of both experiments point to the repetitive variability patterns of the integral $\gamma$--ray flux.
Calculation of the variability index~\citep{Abdo:2010} showed that the overall lightcurve is different from a constant at a $7.1\sigma$ level of significance~\citep{Aliu:2014}.
\par 
In our calculations, we used the data gathered by the VERITAS and HESS collaborations from Table~1 in~\citep{Aliu:2014}.
Data from individual observational epochs do not come from longer time periods than one month.
Fig.~\ref{fig:J0632_MJD} shows confidence intervals for the source parameter~$\beta$ at a $3\sigma$ level of significance together with values $\beta = \on / \alpha \off$ determined for each measurement.
Confidence intervals are ordered chronologically according to the Modified Julian Date of observations.
In Fig.~\ref{fig:J0632_MJD}, also evaluated average source parameters deduced from the HESS (red circles) and VERITAS (blue squares) data taken over complete observational campaigns, $\beta_{\mathrm{HESS}}=1.69$ and $\beta_{\mathrm{VER}}=3.02$, respectively, are shown.
Inspecting the HESS confidence intervals alone, none of the measurements result in extraordinary values of the source parameter with respect to the others.
On the other hand, the joint set of HESS and VERITAS observations provides several non--overlapping pairs of confidence intervals.
Looking at each of the individual measurements, the resultant Li--Ma significances for the test of the null hypothesis given by the average source parameter $\beta = 2.46$ ascertained from the whole set of observations are $-5.8 < \LM < 4.2$.
We can thus conclude that the $\gamma$--ray intensity of the binary system \J~ cannot be considered steady at least on time scales of months in accord with \citep{Aliu:2014}.
\par 
In order to examine the evolution of the source activity due to the orbital motion of the compact object around the Be--star, we folded the numbers of on-- and off--source counts with the period of 315 days.
The orbital period was derived by~\citet{Aliu:2014} using the results of measurements by the Swift satellite which resolved recurring features in the X--ray lightcurve.
Phase 0 was set as $\mathrm{MJD}_{0}=54857$, corresponding to first Swift observations.

\begin{figure}[t!]
	\begin{center}
		\includegraphics[width=\columnwidth]{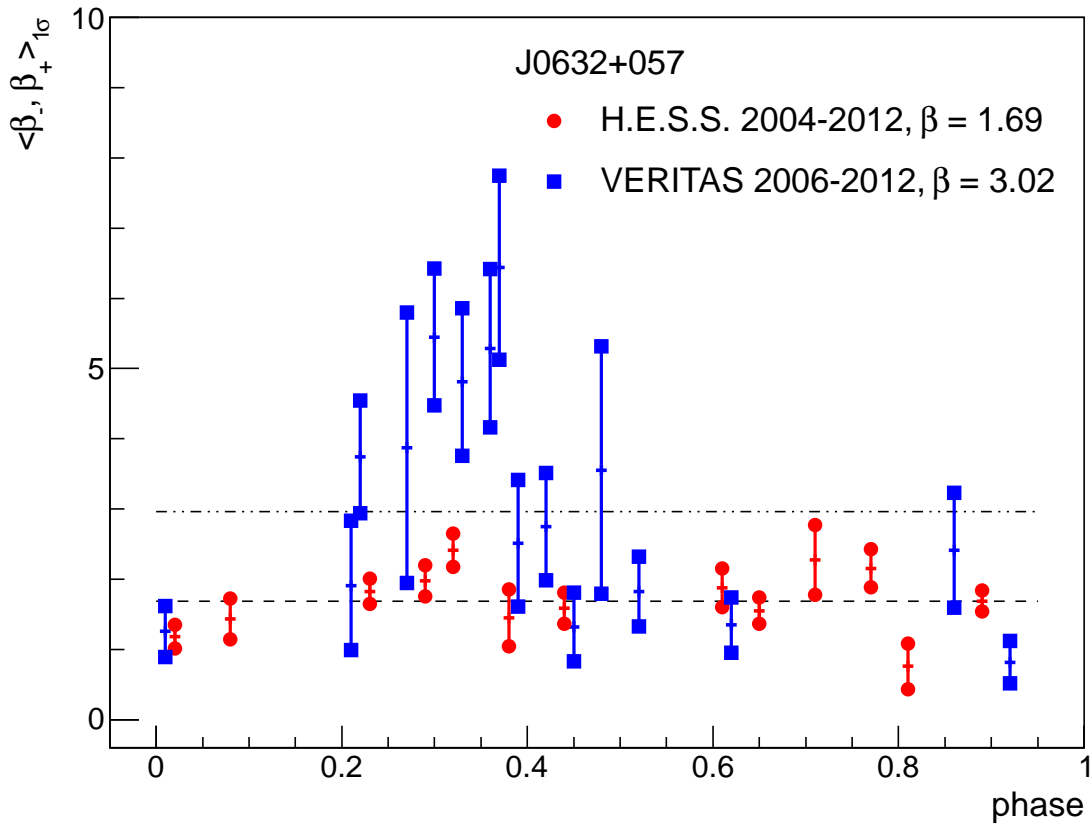}
		\caption{
			$68\%$ confidence intervals for the source parameter $\beta$ of the $\gamma$--ray binary \J~\protect\citep{Aliu:2014} are depicted.
			The measurements are folded with the orbital period of 315 days.
			For further details see caption to Fig.~\ref{fig:J0632_MJD}.
		}
		\label{fig:J0632_phase}
	\end{center}
\end{figure}

\par 
Phase--ordered confidence intervals at a $1\sigma$ level of significance are depicted in Fig.~\ref{fig:J0632_phase}.
The choice of the level of significance in Fig.~\ref{fig:J0632_phase} was motivated by the comparison with the similar plot for fluxes in~\citep{Aliu:2014} where $1\sigma$ statistical uncertainties were depicted.
Several confidence intervals obtained from VERITAS data around phases $\sim0.3$ do not overlap with the remaining ones.
This distinctive rise in the \J~$\gamma$--ray activity correlates with an increase in intensity visible in the sequence of confidence intervals obtained using the HESS data.
Possibly enhanced emission might be emerging at phases $\sim0.7-0.9$.
\par 
The phase folded X--ray flux measured by Swift exhibits a double--peak structure with a prominent maximum around phases $\sim0.35$ and a secondary peak of about half the size of the main one at phases $\sim 0.6-0.9$~\citep{Aliu:2014}.
Besides these periods of increased emission, a clear dip can be recognized at phases $\sim0.45$ suggesting that either an absorption process or a termination of particle acceleration is taking place during the particular arrangement of the massive object and its companion star~\citep{Aliu:2014}.
Our analysis of confidence intervals for the $\gamma$--ray intensity reveals a presence of a peak consistent with the one observed at phases $\sim0.35$ in the X--ray band.
In agreement with the analysis of variability of the integral flux with $1\sigma$ uncertainties done by the VERITAS and HESS collaborations~\citep{Aliu:2014}, hints of increased VHE activity around the position of the second X--ray peak arise in our approach for the 68\% confidence intervals.
However, we did not find any significant inconsistency in the sequence of confidence intervals at a $3\sigma$ level of significance deduced from the HESS observations, which cover the inspected range of phases.
Neither did we reveal any signs of absorption of the $\gamma$--rays.


\section{Conclusions}

Temporal changes in the observed $\gamma$--ray intensities of the binary \J~ we\-re studied by the means of the modified on--off method with the emphasis put on its usefulness in the analyses of data gathered by Cherenkov telescopes.
The assumption of the constant source activity was ruled out consistently with the previous findings on time variability of the binary system~\citep{Aliu:2014}.
The correlation of the VHE $\gamma$--ray activity with the observations in the X--ray energy band was verified at least in terms of the regularity of emission and increased activity at particular phases.
\par
The use of the modified on--off scheme is justified by a strong statistical motivation as well as its straightforward application, while remaining sufficiently general at the same time.
Absence of any need for complex calculation of fluxes in the modified on--off method makes it a convenient choice for investigation of observed intensity variations in VHE $\gamma$--astronomy.


\section*{Acknowledgements}
This work was supported by the Czech Science Foundation grant 14-17501S.



\end{document}